\documentclass{article}
\usepackage{graphicx,picinpar}

\usepackage{color}

\usepackage{cite}
\usepackage{textcomp}
\usepackage{amsmath,amsfonts,amssymb}
\usepackage{enumitem}
\usepackage[utf8]{inputenc}
\usepackage{caption}
\usepackage[normalem]{ulem}
\usepackage{soul}

\newcommand{\pval}{\text{p-value}}

\usepackage{tikz}
\usetikzlibrary{calc,arrows,decorations.pathmorphing,intersections}

\begin{document}

\title{Detection limit of a lutetium based non-paralizable PET-like detector.}

\author{Emmanuel~Busato\\ Laboratoire de Physique de Clermont (Particules, pLasmas, Univers, applicationS), \\ Université  Clermont Auvergne, CNRS/IN2P3, F-63000 Clermont-Ferrand, France\\
  and\\
\'Edouard Roux}

\maketitle

\begin{abstract}

The effect of the intrinsic lutetium radioactivity on the detection performances of a LYSO based \textit{in-beam} PET-like prototype used for quality control of hadrontherapy treatments is studied.
This radioactivity leads to a background that degrades the measurement of the $\beta^+$ signal.
In particular, it prevents the measurement of faint signals originating from low activity $\beta^+$ sources.
This paper presents a method to estimate the minimum $\beta^+$ activity that can be measured for any acquisition time taking into account the non-extensible dead time of the detector.
This method is illustrated with experimental data collected with the \textit{in-beam} PET-like prototype.
The results presented in this paper are therefore specific to this detector.
The method can however be applied in other contexts, either to other lutetium based PET detectors or even to non-PET detectors affected by lutetium radioactivity.
The dead time correction formalism can also be used generally to scale signal and background yields in any non-paralizable detector, even those in which the background is not due to the presence of intrinsic radioactivity.

\end{abstract}

\section{Introduction}

Detectors using LYSO or LSO scintillators as active material are widely used nowadays, in particular for positron emission tomography (PET) measurements. 
Such materials contain a mixture of two lutetium isotopes: $^{175}$Lu, which is stable, and $^{176}$Lu, which is a $\beta^-$ emitter.
The natural abundance of $^{176}$Lu is equal to about 2.6\%. 
This value typically leads to a $\beta^-$ activity of about 300 Bq/cm$^3$ in LYSO or LSO crystals.
This gives rise to an intrinsic and irreducible background (hereafter denoted as lutetium background or simply background) that often needs to be accounted for in the measurement.

PET signal events are made of two back-to-back 511 keV photons coming from the $\beta^+$ annihilation in the patient. 
The lutetium background can lead to energy depositions in the detector similar to those of the signal.
It therefore contributes to the counting, polutes analysis samples and limits the detector sensitivity to signal events.

In this paper, the effect of the lutetium background in non-paralizable PET-like detectors\footnote{A non-paralizable detector is a detector with a non-extensible dead time.} is studied. 
A method to quantitatively estimate the detector detection limit, where detection limit is defined as the minimum $\beta^+$ activity the system can detect for a given acquisition time, is described.
This method accounts for the non-extensible dead time of the detector. 
It is illustrated using experimental data collected with an \textit{in-beam} PET-like prototype (called LAPD for Large Area Pixelized Detector) used for hadrontherapy beam ballistic control.

It should be stressed that only the detection limit problematic is addressed in this paper. 
The only type of assessment the proposed method allows to make is whether or not the detector is able to detect the signal, not whether the detector is able to perform the physics measurements it was initially built for (such as reconstructing the $\beta^+$ activity distribution).
Evaluating the detector ability to perform the physics measurement it was built for requires more studies and is beyond the scope of this paper.

Even if this method is applied to a detector used for hadrontherapy PET-like measurements it can be used with other detectors and in other contexts. 
It can for example be applied to classical PET detectors where the aim is to measure the $\beta^+$ activity coming from the injection in the patient of $\beta^+$ radionuclide.
It can also be applied to 
more generic experimental measurements as long as a signal 
strength parameter
analog to the $\beta^+$ activity has been identified (such as the signal production cross-section in collision events for example) and a technique to estimate the background in the signal region has been found.

The paper is organized as follows.
The experimental setup, data and event selection are described in Sec.~\ref{Sec:ExperimentalSetup}. 
Some details about the lutetium background are given in Sec.~\ref{sec:LuBackground}. 
The method used for calculating the detection limit is described in Sec.~\ref{Sec:SensitivityCalc}.
Results are described in Sec.~\ref{Sec:Results}.

\section{Experimental setup, data and event selection}
\label{Sec:ExperimentalSetup}

\subsection{Detector}

Only the characteristics of the LAPD detector relevant for the studies presented in this paper are described.
For a more complete description, the reader is referred to \cite{8168281}.

The LAPD detector is made of two identical sides constituted of 120 channels each. 
Each channel is made of one LYSO crystal with dimensions $13\times 13\times 15$~cm$^3$ coupled to one photomultiplier tube.
The signal produced by each channel is sent to the custom-made ASM (Analog Sampling Module) electronic boards.
There is a total of 12 ASM boards in the system (each of them treats 20 channels).
Each ASM board performs two main tasks. 
Firstly, it generates trigger signals that are sent to a trigger board for trigger decision. 
Secondly, once a positive trigger decision is taken by the trigger board, each ASM board digitizes channels' datas and transfers them to a CPU board via the VME backplane.
The CPU board filters data, rejecting channels for which the signal is below some threshold, and sends them to a PC for monitoring and analysis.
The transfer between the ASM boards and the CPU board induces a dead time during which no positive trigger decision can be taken.
This dead time is always the same, non-extensible and equal to 41~ms.
Accounting for this non-extensible dead time is one of the main feature of the method described in Sec.~\ref{Sec:SensitivityCalc}.

\subsection{Data}
\label{Sec:Data}

The data used in this paper were obtained in two runs: 
\begin{itemize}
\item \textit{Background run}: no $\beta^+$ source was present in the field of view of the detector. 
The only physical process contributing to the count rate and thus to the data sample is the lutetium background coming from the $\beta^-$ decay of $^{176}$Lu.
This background is discussed in more details in Sec.~\ref{sec:LuBackground}.
The number of events collected with the LAPD for this run is 500,000.
\item \textit{Signal run}: a $\beta^+$ emitting $^{22}$Na source was placed at the center of the field of view of the detector. 
The activity of the source is equal to $16$~kBq. 
As only $90.4$\% of decays lead to the emission of a positron, the $\beta^+$ activity is $14.4$~kBq.
The number of events collected with the LAPD for this run is 14,843, corresponding to an acquisition time of $10.9$~min.
\end{itemize}

All detector, trigger and data acquisition parameters are exactly the same for both runs.

It should be noted that, in the \textit{signal run}, the $^{22}$Na source was not placed inside any tissue equivalent material.
The main material surrounding the source was air.
This minimizes the effect of diffusion and may lead to better detection limits than the ones expected in clinical conditions.

\subsection{Event selection}
\label{Sec:EventSelection}

Two levels of selection are considered. 
The first one is the online selection performed by the trigger system.
A positive trigger decision is taken when at least two channels (one on each side of the LAPD) 
record a pulse with an energy between approximately $250$~keV and $1000$~keV and the time difference between two of these pulses is lower than 20 ns. 
The second one is an offline selection applied to all events selected by the trigger system.
It selects events which have exactly two pulses with an energy $E\in\left[421;601\right]$~keV and a reference
time difference in absolute value lower than $3.95$~ns (the reference time is defined as the time for which the amplitude on the rising front of the pulse reaches $30$~\% of its maximum value).
These criteria correspond both to a $3\sigma$ window around the expected values for PET signal events (the expected values being $511$~keV for the pulses' energies and $0$ for the time difference). 
This second selection is called the "signal selection" and events passing it are said to belong to the "signal region".

\section{Intrinsic lutetium background}
\label{sec:LuBackground}

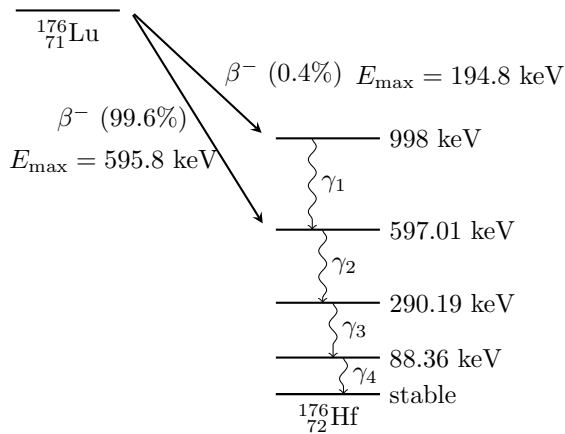
\begin{figure}[!htb]\centering
  \captionsetup{justification=centering}
  \resizebox{7cm}{!}{
    \begin{tikzpicture}[
      scale=0.7,
      level/.style={thick},
      virtual/.style={thick,densely dashed},
      trans/.style={thick,->,shorten >=2pt,shorten <=2pt,>=stealth},
      classical/.style={thin,double,<->,shorten >=4pt,shorten <=4pt,>=stealth},
      radiative/.style={->,decorate,decoration={snake,amplitude=1.5}}
    ]
    \draw[level] (0cm,13em) -- (2cm,13em) node[midway,below] {$^{176}_{~71}$Lu};
    \draw[level] (5cm,6em) -- (7cm,6em) node[right] {$998$~keV};
    \draw[level] (5cm,1em) -- (7cm,1em) node[right] {$597.01$~keV};
    \draw[level] (5cm,-3em) -- (7cm,-3em) node[right] {$290.19$~keV};
    \draw[level] (5cm,-6em) -- (7cm,-6em) node[right] {$88.36$~keV};
    \draw[level] (5cm,-8em) -- (7cm,-8em) node[right] {stable};
    \draw[level] (5cm,-8em) -- (7cm,-8em) node[midway, below] {$^{176}_{~72}$Hf};
    \draw[trans] (2.2cm,13em) -- (4.8cm,6.1em) node[midway,right] {$~~\beta^-$ (0.4\%)};
    \draw[trans] (2.2cm,13em) -- (4.8cm,1.1em) node[midway,left] {$\beta^-$ (99.6\%)};
    \draw[radiative] (5.7cm,6em) -- (5.7cm,1em) node[midway,right] {$\gamma_1$};
    \draw[radiative] (5.9cm,1em) -- (5.9cm,-3em) node[midway,right] {$\gamma_2$};
    \draw[radiative] (6.1cm,-3em) -- (6.1cm,-6em) node[midway,right] {$\gamma_3$};
    \draw[radiative] (6.3cm,-6em) -- (6.3cm,-8em) node[midway,right] {$\gamma_4$};

    \end{tikzpicture}
  }
\put(-198,103){$E_{\text{max}}=595.8~$keV}
\put(-67,135){$E_{\text{max}}=194.8~$keV}
\caption{Decay scheme of $^{176}_{~71}$Lu~\cite{KOSSERT2013140}. \label{Fig:LuDecayScheme}} 
\end{figure}

$^{176}$Lu is a radioactive isotope with a half-life of $3.6\times~10^6$~years. Its decay scheme is shown in Fig.~\ref{Fig:LuDecayScheme}. 

Most of the time the $^{176}$Lu isotope undergoes $\beta^-$ decay by emission of an electron with a maximum energy of $595.8$~keV. 
This $\beta^-$ decay is followed by the emission of three photons of energy $306.82$, $201.83$ and $88.36$~keV.
This can lead to two different types of events in the LAPD.
The first one (hereafter denoted as type-I events) corresponds to the random coincidence between two $\beta^-$ decays on each side of the detector. 
After the $\beta^-$ decay, the electron deposits an energy between $0$ and $595.8$~keV in a LYSO crystal. 
The photons can also deposit part of or all their energy in the same crystal. 
All these energy depositions can lead, in one crystal, to an energy comprised in the trigger energy window defined in Sec.~\ref{Sec:EventSelection}. 
If two such lutetium $\beta^-$ decays occur on each side of the LAPD within $20$~ns a positive trigger is taken.

The second type of events (hereafter denoted as type-II events) corresponds to a $\beta^-$ decay in a crystal on one side of the LAPD and to the interaction of the subsequent $306.82$~keV photon in another crystal on the other side of the detector.
Type-II events correspond to a true coincidence because the elapsed time between the $\beta^-$ decay and the $306.82$~keV photon emission and the time it takes for the photon to travel from one side of the detector to the other are lower than the time resolution of the detector. This type of events can also lead to positive trigger decisions as both the $\beta^-$ and the $306.82$~keV photon energy depositions can be comprised in the trigger energy window.

\begin{figure}[h] \centering
\includegraphics[width=8cm]{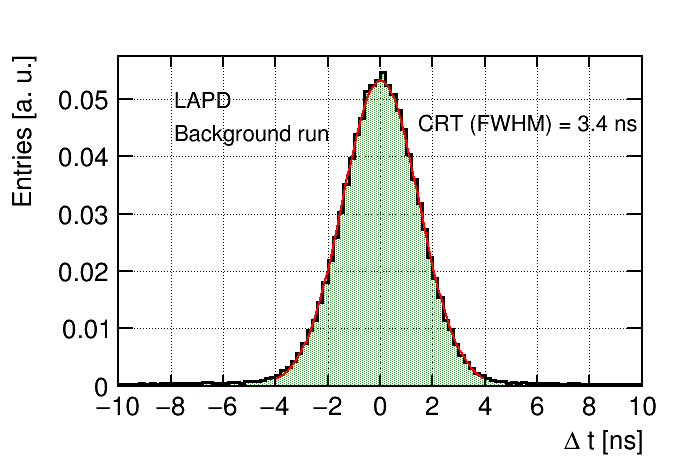} 
\caption{Distribution of the time difference between the two pulses after trigger selection in the \textit{background run}. A gaussian function fit to the data is shown in red. Only events with exactly two pulses are considered.\label{Fig:CRT}}
\end{figure}

The proportion in which type-I and type-II events contribute to data samples depends on the applied event selection. 
After trigger selection, it is found that samples are enriched in type-II events.
This can be seen from the distribution of the reference time difference between the two pulses in the \textit{background run} data shown in Fig.~\ref{Fig:CRT}.
This distribution shows that most pairs of pulses have a time difference which is normally distributed (as one would expect for coincidence events) and not flat (as one would expect for random coincidences). 

After signal selection, only type-I events are selected. 
\mbox{Type-II} events are rejected because $306.82$~keV photons cannot deposit a sufficient energy to be selected in the signal region.
Type-I events are irreducible, as opposed to type-II events. 
Even by tightening the signal selection (by applying for example narrower energy and time difference selection windows), there will always be a certain fraction of type-I events passing it.
With the current event selection criteria, both type-I and type-II events limit the detector sensitivity to signal events.
By tightening the energy window in the trigger selection it would in principle be possible to reject type-II events directly at trigger level and thus to decrease the detection limit.
This would however, at least with the detector and method presented in this paper, complexify the estimation of the background yield in the signal region. 
Indeed, as will be seen in Sec.~\ref{Sec:DeterminationYieldsNominalActivity}, events passing the trigger selection but not the signal selection are used to normalize the background yield in the signal region. 

\section{Detection limit calculation}
\label{Sec:SensitivityCalc}

The purpose of this work is to estimate the system's detection limit, where detection limit is defined as the signal detection threshold in terms of activity. 
The method used to perform this estimation is described in Sec.~\ref{Sec:Method}. 
Subsequent sections are dedicated to the description of the calculation of the various quantities needed by the method.

\subsection{Method}
\label{Sec:Method}

The goal is to find the minimum activity $\beta^+$ samples must have in order to observe the signal on top of the lutetium background for any given acquisition time. 
Finding a solution to this problem requires to precisely define what is meant by observing a signal in the presence of background and to account for an important characteristic of non-paralizable detectors which is that there is a non-extensible dead time contributing to event loss. 
The method proceeds in two steps:

\begin{itemize}
\item Step 1: calculation of the signal and background yields as a function of the $\beta^+$ activity for a given acquisition time, 
\item Step 2: calculation of the signal observation significance $S$ as a function of the $\beta^+$ activity and determination of the activity for which $S$ is equal to some predefined value. 
In the following, the value $S=3$ will be used.
This value is generally used in particle physics as the threshold above which a signal is said to be observed.
\end{itemize}

Step 1 could in principle be performed using several $\beta^+$ radioactive sources with different activities. 
However, the required number of sources would typically be very large and it seems impossible to perform such measurements in practice.
Another approach is thus used, itself subdivided into two steps: 
\begin{itemize}
\item Step 1.a: determination of initial signal and background yields from a measurement with a $\beta^+$ source with known activity $a_1$, 
\item Step 1.b: scaling of the initial signal and background yields with the $\beta^+$ activity. 
\end{itemize}

This approach needs only one $\beta^+$ source. 
Data collected with this unique source can then be used to determine signal and background yields for any other activity, using a formalism that accounts for the non-extensible dead time of the detector.

In the rest of this paper, the method is illustrated using the data described in Sec.~\ref{Sec:Data}.
As the $^{22}$Na source was placed at the center of the field of view for the \textit{signal run}, the detection limit computed is that at the center of the field of view.
Calculating the detection limit at any other position in the field of view can be done using exactly the same method by moving the source to the desired position. 

\subsection{Determination of initial yields (step 1.a)}
\label{Sec:DeterminationYieldsNominalActivity}

\begin{figure}[!htb] \centering
\includegraphics[width=9cm]{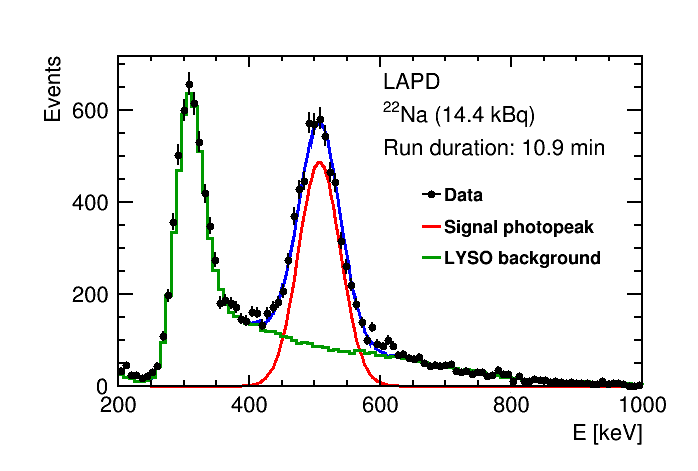} 
\caption{Energy spectrum measured with the LAPD detector in the \textit{signal run} with a 14.4 kBq $^{22}$Na source. The acquisition time is 10.9~minutes. The lutetium background is shown in green and the 511~keV signal photopeak is shown in red. \label{Fig:EnergySpectrumFit}}
\end{figure}

In order to determine initial signal and background yields, the data acquired in the \textit{signal} and \textit{background runs} described in Sec.~\ref{Sec:Data} are used. 
Yields are determined from a fit to the energy spectrum in the \textit{signal run} (see Fig.~\ref{Fig:EnergySpectrumFit}).
The energy variable is used because it shows good discriminating power between signal and background events but any other discriminating variable could be used instead.

The fit was performed in the range $E\in\left[650;1000\right]$~keV with the maximum likelihood method.
A binned likelihood has been used.
Only the background spectrum measured in the \textit{background run} was fitted to the data as very few signal events are expected to contribute in this range (the signal photopeak in Fig.~\ref{Fig:EnergySpectrumFit} is shown just for illustrative purpose).
The reason for using this range and not including the energy distribution for signal events in the fit is that this distribution is currently not precisely known.
The signal distribution cannot be simply modeled by a single gaussian with mean equal to $511$~keV as 
some 511~keV photons do not deposit all their energy but only part of it due to compton interaction. 
The energy spectrum for signal events is therefore expected to have, in addition to the 511~keV peak, features specific of the compton interaction such as the compton continuum and the compton edge.
In order to precisely know the signal energy distribution, a full simulation of the LAPD detector can be used.
However, at the time of writing this paper, this simulation was not finished and only the background spectrum in a background enriched region was used.

The fitted background yield is equal to 9415.
The signal yield, given by the difference between the observed number of events and the background yield, is equal to 5428.
It should be noted that the exact approach used to calculate the initial signal and background yields is not important. 
The rest of the method only needs the value of these initial yields, nomatter how they are calculated.

It is also important to note that all events passing the trigger must be considered in the determination of the initial signal and background yields (Fig.~\ref{Fig:EnergySpectrumFit} is obtained without any selection besides trigger selection). 
The reason is that the scaling formalism described in Sec.~\ref{Sec:YieldsScaling} would otherwise be invalid.

\subsection{Scaling of yields with activity (step 1.b)}
\label{Sec:YieldsScaling}

In the following, the $\beta^+$ activity will be written as 
\begin{equation}
a = \alpha a_1
\end{equation} 
where $a_1$ is the activity of the source used in step 1.a (see Sec.~\ref{Sec:Method}) and $\alpha$ is the scaling parameter.

For detectors and data acquisition systems with no dead time (or with a dead time too small compared to the inverse counting rate such that it does not lead to event loss), the scaling of the initial yields with the $\beta^+$ activity ($a$) is straightforward. 
In this case:
\begin{equation}
\label{Eq:ScalingBackIdealCase}
m_b^{\alpha} = m_b^{1}
\end{equation}
\begin{equation}
\label{Eq:ScalingSigIdealCase}
m_s^{\alpha} = \alpha\times m_s^{1}
\end{equation}
where $m_b^{1}$ and $m_s^{1}$ are respectively the initial background and signal rates determined in Sec.~\ref{Sec:DeterminationYieldsNominalActivity} and $m_b^{\alpha}$ and $m_s^{\alpha}$ are the scaled rates.
Yields are then derived from these rates by multiplication with the acquisition time.
The equality in Eq.~\ref{Eq:ScalingBackIdealCase} indicates that, for a given acquisition time, the intrinsic lutetium background yield is constant.

In the LAPD system, a non-extensible dead time contributing significantly to event loss is present. 
Eq.~\ref{Eq:ScalingSigIdealCase} and \ref{Eq:ScalingBackIdealCase} thus cannot be used. 
Instead, a proper formalism accounting for the dead time must be used.

For non-extensible dead time and in cases where only one physical process contributes to the measurement, it is well-known that the true rate $r$ and measured rate $m$ are related by the following equation~\cite{Knoll:1300754}:
\begin{equation}
\label{Eq:DeadTimeEquationBasic}
m = \frac{r}{1+r\tau}
\end{equation}
where $\tau$ is the dead time.

In cases where multiple physical processes contribute to the measurement Eq.~\ref{Eq:DeadTimeEquationBasic} remains valid but $r$ and $m$ become respectively the sum of the true and measured rates over all processes:
\begin{equation}
r = \sum\limits_{p=1}^P r_p
\end{equation}
and
\begin{equation}
m = \sum\limits_{p=1}^P m_p
\end{equation}
where $p$ denotes the process and $P$ is the total number of processes.
The measured rate for a given process $p$ is given by
\begin{equation}
\label{Eq:mpVsrptau}
m_p = \frac{r_p}{1+r\tau}
\end{equation}

In the experiments considered in this paper only two types of processes contribute to the counting rate: the background from lutetium decay and the signal from $\beta^+$ decay.
As described in Sec.~\ref{sec:LuBackground}, background events from lutetium decay can in fact originate from two different processes. 
It is therefore tempting to consider not only one type of process for the lutetium background but two. 
Even if this would of course be possible, the different types of background processes are considered a single type of process because 
the fit performed in Sec.~\ref{Sec:DeterminationYieldsNominalActivity} does not allow to distinguish them and only gives the total background yield.
Background and signal rates are thus given by:
\begin{equation}
\label{Eq:mpVsrptauForSignalAndBackground}
m_b^{\alpha} = \frac{r_b}{1+(r_b + r_s^{\alpha})\tau}~\text{and}~m_s^{\alpha} = \frac{r_s^{\alpha}}{1+(r_b + r_s^{\alpha})\tau}
\end{equation}
with
\begin{equation}
\label{Eq:rsAlphaVsAlpha}
r_s^{\alpha}=\alpha\times r_s^1
\end{equation}
where $r_s^1$ is the true signal rate corresponding to the activity $a_1$.
From Eq.~\ref{Eq:mpVsrptauForSignalAndBackground} and Eq.~\ref{Eq:rsAlphaVsAlpha} it is possible to express the measured yields as a function of $\alpha$. 
This requires however the prior knowledge of $r_s^1$ and $r_b$.
In order to determine these values, the measurement performed for $\alpha=1$ in Sec.~\ref{Sec:DeterminationYieldsNominalActivity} can be used.
For $\alpha=1$, Eq.~\ref{Eq:mpVsrptauForSignalAndBackground} and \ref{Eq:rsAlphaVsAlpha} give:
\begin{equation}
\label{Eq:mpVsrptauForSignalAndBackgroundAlphaEq1}
m_b^{1} = \frac{r_b}{1+(r_b + r_s^{1})\tau}~\text{and}~m_s^{1} = \frac{r_s^{1}}{1+(r_b + r_s^{1})\tau}
\end{equation}

These two equations can be solved for $r_b$ and $r_s^1$ which can then be inserted in Eq.~\ref{Eq:rsAlphaVsAlpha} and Eq.~\ref{Eq:mpVsrptauForSignalAndBackground}. 
Equivalently, Eq.~\ref{Eq:mpVsrptauForSignalAndBackground}, Eq.~\ref{Eq:rsAlphaVsAlpha} and Eq.~\ref{Eq:mpVsrptauForSignalAndBackgroundAlphaEq1} can be used to express scaled yields as a function of initial yields as follows:
\begin{equation}
\label{Eq:FinalScalingEqBack}
m_b^{\alpha} = \frac{m_b^{1}}{1+m_s^{1}\tau\left(\alpha-1\right)}
\end{equation}
\begin{equation}
\label{Eq:FinalScalingEqSig}
m_s^{\alpha} = \frac{\alpha m_s^{1}}{1+m_s^{1}\tau\left(\alpha-1\right)}
\end{equation}

\begin{figure}[h] \centering
\includegraphics[width=9cm]{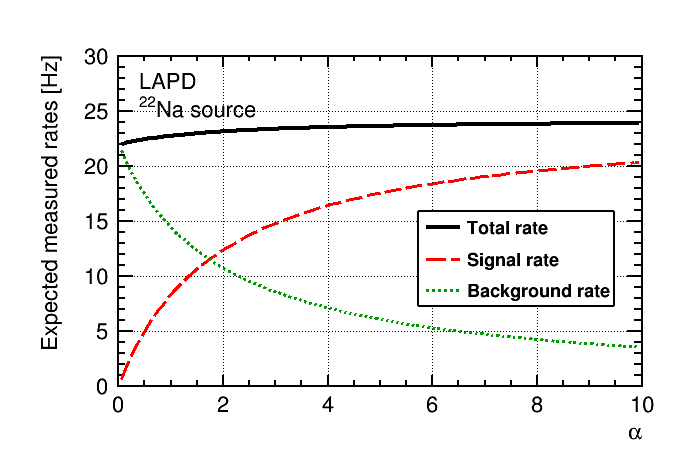} 
\caption{Total, signal and background rates as a function of the activity scaling parameter $\alpha$. \label{Fig:ScalingVsAlphaTotSigBkg}}
\end{figure}

It is straightforward to check that, if $\tau=0$, Eq.~\ref{Eq:FinalScalingEqBack} and \ref{Eq:FinalScalingEqSig} lead back to Eq.~\ref{Eq:ScalingBackIdealCase} and \ref{Eq:ScalingSigIdealCase}.
For $\tau> 0$, both $m_s^{\alpha}$ and $m_b^{\alpha}$ depend on $\alpha$ while only $m_s^{\alpha}$ depends on it when $\tau=0$. 
Figure~\ref{Fig:ScalingVsAlphaTotSigBkg} shows $m_s^{\alpha}$ and $m_b^{\alpha}$ as a function of $\alpha$ with $m_s^{1}$ and $m_b^{1}$ equal to the values found in Sec.~\ref{Sec:DeterminationYieldsNominalActivity} and with $\tau=41$~ms.

As already mentionned at the end of Sec.~\ref{Sec:DeterminationYieldsNominalActivity}, it is important to note that, in Eq.~\ref{Eq:FinalScalingEqBack} and \ref{Eq:FinalScalingEqSig}, background and signal rates are rates after trigger selection. 
These equations should not be used to scale rates after signal selection.
This is because the dead time is consecutive to a positive trigger decision. 
The amount of events lost due to dead time and their scaling with the activity is therefore governed by the frequency at which positive trigger decisions are taken.

From Eq.~\ref{Eq:FinalScalingEqBack} and Eq.~\ref{Eq:FinalScalingEqSig}, it is possible to calculate the background and signal rates (and the corresponding yields) for any activity from the initial rates $m_s^{1}$ and $m_b^{1}$ and thus to derive the detection limit of the system, as long as the dead time $\tau$ is known.

\subsection{Calculation of the observation significance (step 2)}

The significance $S$ of an observation is a measure of the agreement between the observed number of events and the background only hypothesis. 
As the value of the significance increases, the observation is less and less compatible with the background only hypothesis and it is more and more likely that some signal contributes to the observation. 
It is customary to consider that the observation of the signal is statistically significant when the significance is equal to $3$ or greater.
The detection limit is therefore defined as the activity for which $S=3$. 

In order to compute the significance of the observation as a function of the activity, a frequentist statistical hypothesis test is performed for several values of $\alpha$. 
The statistical test requires the statistical model to which the measurement obeys to be determined. 
For the measurements considered in this paper, the observed number of events $N$ is a Poisson random variable.
The model can thus be written as follows:
\begin{equation}
\label{Eq:StatisticalModel}
P\left(N; s, b\right) = \frac{\left(s+b\right)^{N}}{N!}e^{-\left(s+b\right)}
\end{equation}
where $s$ and $b$ are the signal and background yields respectively.
They are not simply the yields obtained from the rates in Eq.~\ref{Eq:FinalScalingEqBack} and Eq.~\ref{Eq:FinalScalingEqSig} but the yields in the signal region defined in Sec.~\ref{Sec:EventSelection}.
The reason for applying the signal selection is to improve the signal to background ratio with respect to that after trigger selection (that one would have by using Eq.~\ref{Eq:FinalScalingEqBack} and Eq.~\ref{Eq:FinalScalingEqSig} directly).
$s$ and $b$ can thus be written as 
\begin{equation}
\label{Eq:sAndbAfterSignalSelection}
s = \varepsilon_s\times m_s^{\alpha}\times t\quad\text{and}\quad
b = \varepsilon_b\times m_b^{\alpha}\times t
\end{equation}
where $t$ is the acquisition time and $\varepsilon_s$ and $\varepsilon_b$ are respectively the signal and background efficiencies (probabilities for signal and background events to pass the signal selection).

The background efficiency is computed by applying the signal selection to the data from the \textit{background run} and by calculating the ratio between the number of events that pass the selection and the total number of events.
It is found to be $0.25$\%.

The signal efficiency is computed using the data from the \textit{signal run}.
The signal selection is applied and the number of events that pass the selection $N_{SR}$ is counted.
The signal efficiency is given by
\begin{equation}
\varepsilon_s = \frac{N_{SR} - \varepsilon_b\times N_b^1}{N_s^1}
\end{equation}
where $N_b^1$ and $N_s^1$ are the initial yields computed in Sec.~\ref{Sec:DeterminationYieldsNominalActivity}. The signal efficiency is found to be $65.5$\%.

Computing the significance from Eq.~\ref{Eq:StatisticalModel} can be done in many ways, in particular in the presence of systematic uncertainties. 
In this paper the purpose is to illustrate the detection limit calculation method rather than to provide a complete review on significance calculation techniques. 
The technique described below is therefore a simple analytical technique not taking into account systematic uncertainties. 
When systematic uncertainties are large, users should prefer using other techniques.
In the case where only one source of systematic uncertainty affects the background yield users can for example use the technique described in~\cite{Choudalakis2012}.
In the case of multiple sources of systematic uncertainties (with potential correlation between them), users can for example use numerical tools such as OpTHyLiC~\cite{Busato:2015ola} or RooStats~\cite{roostats}.
These tools are more difficult to use in practice than the simple analytical technique presented below but lead to more accurate results when the effect of systematic uncertainties is non-negligible.

When $s$ and $b$ are perfectly known, the significance is computed from the \pval~of the observation $N_{\text{obs}}$ under the background hypothesis:
\begin{equation}
\label{Eq:pValueGeneral}
\pval = \sum\limits_{N=N_{\text{obs}}}^\infty P\left(N; s=0, b\right)
\end{equation}

In Eq.~\ref{Eq:pValueGeneral}, the observed yield $N_\text{obs}$ is taken, for simplicity, to be equal to the average observed yield under signal plus background hypothesis, that is 
\begin{equation}
\label{Eq:ObservedYieldForPValue}
N_{\text{obs}}=s+b
\end{equation}

The significance $S$ is finally given by 
\begin{equation}
\label{Eq:SignifAsAFunctionOfPVal}
S=\Phi^{-1}\left(1-\pval\right)
\end{equation}
where $\Phi$ is the cumulative distribution function of the standard normal distribution.

Numerical computation of significances with Eq.~\ref{Eq:pValueGeneral}, \ref{Eq:ObservedYieldForPValue} and \ref{Eq:SignifAsAFunctionOfPVal} can be done very easily by using the relation between the poisson cumulative distribution function and the gamma cumulative distribution function:
\begin{equation}
\label{Eq:RelationPoissonGammaCdfs}
\sum\limits_{N=0}^{s+b}P(N; s=0, b)=1-F_\Gamma(b; s+b+1)
\end{equation}
where $F_\Gamma(b; s+b+1)$ is the gamma cumulative distribution function:
\begin{equation}
F_\Gamma(b; s+b+1) = \frac{\displaystyle\int_{0}^{b} x^{s+b}e^{-x}\text{d}x}{\displaystyle\int_{0}^{\infty} x^{s+b}e^{-x}\text{d}x}
\end{equation} 
For example, with the ROOT software~\cite{ANTCHEVA20092499}, the significance is computed with a one line command:

\begin{center}
\small \verb|double S = ROOT::Math::normal_quantile(1-|\verb|ROOT::Math::gamma_cdf(b,s+b, 1), 1);|
\end{center}

\section{Results}
\label{Sec:Results}

\begin{figure}[!htb] \centering
\includegraphics[width=9cm]{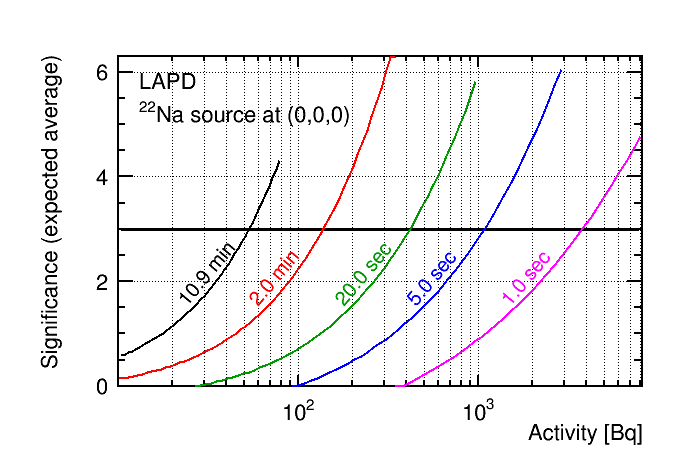} 
\caption{Expected average significance as a function of the activity for different acquisition times. The intersections between the curves and the $S=3$ value give the detection limit for the various acquisition times.\label{Fig:ResultSignifVsActivity}}
\end{figure}

Even with the simple significance calculation method presented in the above section, it is not possible to solve the equation $S=3$ (with $S$ given by Eq.~\ref{Eq:SignifAsAFunctionOfPVal}) for the activity analytically. 
It is therefore necessary to perform an activity scan and to search the value for which $S=3$.
Examples of scans for various acquisition times are presented in Fig.~\ref{Fig:ResultSignifVsActivity}.

The black (leftmost) curve in Fig.~\ref{Fig:ResultSignifVsActivity} shows results for the acquisition time that was used for the \textit{signal run}, i.e. $10.9$~minutes.
The minimum $\beta^+$ activity that can be detected during that time is $53$~Bq.
In other words, any source with an activity of at least $53$~Bq can be detected in a time of at most $10.9$~minutes.
Sources with a lower activity need a longer acquisition time to be detected.

The LAPD detector considered in this paper is intended to be used for online control of hadrontherapy treatments.
For this application, the signal must be detected with much shorter acquisition times than the one used in the \textit{signal run}.
Ideally, the $\beta^+$ activity distribution measurement should be done a few seconds after the beginning of the treatment.
The signal must therefore be detected in a time of the order of one second.
It is thus necessary to evaluate the detection limit for such acquisition times and to compare it to the expected $\beta^+$ activity induced in the patient by the ion beam during the treatment. 
Results for acquisition times of $1$~second and $5$~seconds are shown in magenta and blue (rightmost curves) in Fig.~\ref{Fig:ResultSignifVsActivity} respectively.
The detection limit is $3780$~Bq for $1$~second and $1078$~Bq for $5$~seconds.
The expected $\beta^+$ activity induced in the patient by the beam can be estimated from the simulation results presented in \cite{0031-9155-45-11-403} for example.
Considering only the $\beta^+$ activity from the $^{15}$O, $^{11}$C and $^{10}$C isotopes and for a proton beam with an energy of $200$~MeV, the number of protons required to produce an activity of $3780$~Bq ($1078$~Bq) is approximately $2\times 10^7$ ($5\times 10^6$).
For beams typically used in protontherapy, these numbers of protons are delivered in a time much shorter than one second.
It is therefore possible to conclude that, in a typical protontherapy treatment and provided that beam induced backgrounds can be rejected (such that only the lutetium background remains), it should be feasible to see the $\beta^+$ signal in a time of the order of $1$ second with the LAPD detector.
The lutetium background in this detector is therefore not a limitation for the observation of the signal in typical \textit{in-beam} PET measurements.

\begin{figure}[!htb] \centering
\includegraphics[width=9cm]{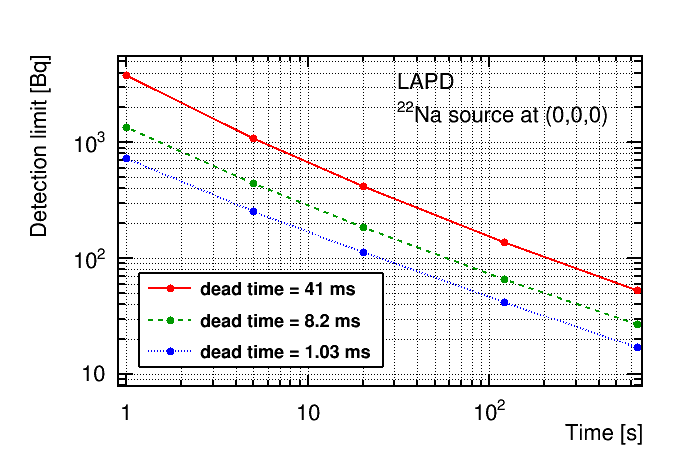} 
\caption{Detection limit as a function of the acquisition time for three values of the detector dead time. \label{Fig:DetectionLimitVsTimeVariousDeadTime}}
\end{figure}

The formalism presented in Sec.~\ref{Sec:YieldsScaling} allows to study the effect of the detection parameters on the detection limit, in particular the dead time. 
Fig.~\ref{Fig:DetectionLimitVsTimeVariousDeadTime} shows the detection limit as a function of the acquisition time for three values of the dead time: $41$~ms, $8.2$~ms and $1.03$~ms. 
The first value, used for the results presented above, is the one of the LAPD detector in its current form.
The second (third) one corresponds to a dead time reduced by a factor of $5$ ($40$) with a high bandwidth data acquisition system.
This figure shows that, for a dead time $40$ times lower than the current one, it requires approximately ten times less time to detect the same $\beta^+$ activity.

\vspace*{-0.5cm}

\section{Conclusion}

The effect of the lutetium induced background in a non-paralizable PET-like detector has been presented.
A simple method to compute the detection limit has been described.
It incorporates a formalism to determine the expected signal and background yields for any activity from data collected with a radioactive source of known activity.
This formalism accounts for event loss due to the non-extensible dead time. 
The method has been illustrated with data collected with an in-beam PET-like detector (LAPD) and a $^{22}$Na source. 

The method can be used with other detectors and in other contexts than the one presented here.
In particular, the yield scaling formalism that accounts for the non-extensible dead time can be used whenever non-paralizable detectors measuring multiple physical processes are used, even when the purpose is not to compute detection limits. 

\section{Acknowledgment}

The authors would like to thank G\'erard Montarou for the careful reading of the manuscript and valuable comments.

\bibliographystyle{abbrv}
\bibliography{Biblio}

\end{document}